\documentclass[aps,prb,twocolumn,showpacs,amsmath,amssymb,
superscriptaddress,10pt]{revtex4-1}

\usepackage{amsmath,amssymb,amsthm}      
\usepackage{amsfonts,bbm,bm}
\usepackage{xcolor}                               
\usepackage{float}
\usepackage[pdftex]{graphicx}\DeclareGraphicsExtensions{.pdf}
\usepackage[colorlinks=true,citecolor=blue]{hyperref}
\usepackage[multidot]{grffile}

\def\up{\mathord{\uparrow}}
\def\down{\mathord{\downarrow}}

\begin{document}
\title{Spin-vibronics in interacting nonmagnetic molecular
nanojunctions} 

\author{S. Weiss}
\affiliation{Theoretische Physik, Universit\"at Duisburg-Essen and CENIDE,
47048 Duisburg, Germany}
\author{J. Br\"uggemann}
\affiliation{I. Institut f\"ur Theoretische Physik, Universit\"at Hamburg,
Jungiusstra{\ss}e 9, 20355 Hamburg, Germany}
\author{M. Thorwart}
\affiliation{I. Institut f\"ur Theoretische Physik, Universit\"at Hamburg,
Jungiusstra{\ss}e 9, 20355 Hamburg, Germany}
\date{\today}
\begin{abstract}
We show that in the presence of ferromagnetic electronic reservoirs and
spin-dependent tunnel couplings, molecular vibrations in nonmagnetic single
molecular transistors induce an effective intramolecular exchange magnetic 
field. It generates a finite spin-accumulation and -precession for the
electrons confined on the molecular bridge and occurs under
(non)equilibrium conditions. The effective 
 exchange magnetic field is calculated here to lowest order in the tunnel
coupling  for
a nonequilibrium transport setup. 
Coulomb interaction between electrons is taken into account as well as
a finite electron-phonon coupling. We show that for realistic
physical parameters, an effective
spin-phonon coupling emerges. It is induced by quantum many-body interactions,
which are either electron-phonon or Coulomb-like. We investigate the 
precession and accumulation of the confined spins as function of bias-
and gate-voltages as well as their dependence on the angle enclosed by the
magnetizations between the left and right reservoir. 
\end{abstract}


\maketitle

\section{Introduction}
Understanding the interplay between Coulomb interaction and the coupling
of the electronic charge to vibrational degrees of freedom has motivated intense
research in
several areas of condensed matter physics in the recent years. Transistors in
future
nanoelectronic applications will ideally work with only a few physical carriers
of information. Inherently, the dimension of the devices are scaled down, such
that 
response times shorten and fast switching between the two
states of a transistor (or a quantum dot) becomes possible. For few electron
quantum dots, the relevant physical properties, such as the number of
confined electrons, effective $g$-factors, and spin-orbit coupling strengths,
are 
coherently controllable by tuning respective gate voltages in experiments, see
Refs.~\onlinecite{Hanson,vdWiel} and references therein. Suitable
physical setups include semiconducting
heterostructures as well as carbon nanotubes \cite{Kuemmeth} or 
gated nanowires \cite{Shorubalko} in order to confine a small number of
electrons. Mechanical degrees of freedom are inherent to the particular geometry
of these devices.

In molecular transistors (or nanojunctions), the electronic and/or spin states
of a molecule are used in a controlled way to realize logical operations
\cite{pederson,molel}. Besides their electronic properties, molecules
inherently possess significant internal vibrational degrees of freedom.
Characteristic frequencies of internal mechanical vibrational modes of a
molecule are used to witness a working device in a transport
setup\cite{Galperin,Tal}. In general, nanomechanical systems designed by
means of a clamped nanobeam, exhibit vibrational degrees of freedom as
well\cite{nems}. By properly adjusting electrostatic gates,
nanomechanical quantum dots with a vibrational
degree of freedom are created. Depending on the particular situation,
the energy scales of the electronic energies and the vibrational motion may
compete and may generate rich physical phenomena.  For example, electronic
energies can be significantly shifted due to a strong electron-vibrational
coupling. This gives rise to the Franck-Condon
blockade as observed in carbon nanotubes
\cite{ensslin}, as well as a strong coupling between the tunneling of electrons
and the mechanical motion \cite{Steele09}. 

In addition to controlling charge states, the electronic spin degree of freedom
has been addressed as well in electronic devices \cite{zutic}. In 
quantum dots, electron spin resonance measurements allow to determine the
particular state of the confined particle. Also in clean nanotubes,
spin-orbit coupling effects serve to further split the single-particle
energy spectrum in the absence of an external magnetic field. There, spin and
valley degrees of freedom may be used to
define appropriate qubits \cite{Kuemmeth, Churchill, Weiss}. Since the
electron spin is a natural two state system, control and coherent
manipulation of a system with a few confined spins is desirable for
applications in spintronics and for encoding quantum information.

As a generic theoretical model for all the mentioned scenarios, the 
 Anderson-Holstein model has been established in the recent years. A 
quantum dot or a molecular nanojunction with a few electronic orbitals and a
finite electron-phonon coupling is tunnel-coupled to large reservoirs of free
electrons. The interplay between the electronic states and the mechanical
motion in Anderson-Holstein setups has been in the focus recently.
Franck-Condon blockade effects together with large Fano factors have been
studied \cite{Koch} as well as the transport properties of the system in the
presence of a finite temperature bias\cite{Leijnse, Koch}. Both cases were
analyzed in the
weak-to-intermediate coupling regime. Thereby, tunneling
processes up to second order in the dot-lead hybridization have been taken into
account. For slow phonons (i.e., small phonon
frequencies),  the adiabatic phonon regime has been
addressed \cite{bode,pistolesi}. Also in this regime, Jovchev {\em et al.\/} 
\cite{Jovchev} have performed a scattering state renormalization group study in
order to compute the $I(V)$ characteristics and to determine the spectral
function under the condition of thermal equilibrium. An advanced perturbative
treatment of the electron-phonon coupling under nonequilibrium conditions has
also been worked out \cite{egger}. Likewise, Monte Carlo simulations have been
performed in the transient regime \cite{Lothar}. Restricting the
Hilbert space to the subspace of a singly occupied
electronic device, Flensberg {\em et al.\/}  \cite{Karsten1,Karsten2} have 
used an incoherent rate equation approach together with the assumption that the
phonon is in thermal equilibrium to reproduce the main 
experimental features on a qualitative level \cite{Park}. The influence on
transport of a nonadiabatic phonon has been studied in Ref.\ 
\onlinecite{Mitra,Koch}. In addition, the crossover between the regime
of the adiabatic and the weak tunneling has been addressed by the iterative
summation of nonequilibrium quantum path integrals in
Ref.~\onlinecite{Huetzen}, suggesting that the Franck-Condon blockade arises 
also at low temperatures. 

Further interesting effects come into play when quantum dots are contacted by
ferromagnetic leads. Then, the reservoirs can induce effective exchange magnetic
fields for the electrons in the quantum dot. This has become known as 
the spin-valve effect. Here, the electronic spin quantum number is no longer 
a good quantum number, and the confined spins are allowed to precess coherently
\cite{Balents, Koenig2, Koenig3}. 
Ferromagnets may be characterized by their difference in the density of
states of the majority and the minority spins.  When two reservoirs are
present (say, the left and the right one) with different
magnetizations, the tunneling magneto-resistance (TMR) depends sensitively on
the angle enclosed by the directions of the two magnetizations. Certainly,
this opens up the possibility to manipulate spin states and initialize or
read out the particular states. A proper theoretical
description has to include a description of the quantum coherences appearing in
the
system. In the present context, quantum coherences arise due to the coherent
spin evolution. A detailed study of the coherent time evolution of the electron
spin  in the presence of an electron-phonon coupling has not been reported so
far in the literature. The influence of the electron-phonon coupling in quantum
transport setups with ferromagnetic leads has been studied by 
means of nonequilibrium Green's function for the two limiting cases of
vanishing Coulomb interaction ($U=0$) and infinitely large Coulomb interaction  
($U\to\infty$) in Refs.~\onlinecite{Cheng,Wang}, respectively. Also the 
case of a single coupled ferromagnet used as a switch for the current
polarization has been investigated recently \cite{Guo}. A Hartree-Fock-like
approximation is used to close the set of equations for the 
causal electronic Green's functions, yielding the tunneling current and the
tunneling magneto-resistance \cite{Rudzinski}. We note that within these
approximations, a quantum coherent spin evolution on the quantum dot is not
obtained due to the absence of inhomogeneous magnetic fields and/or
approximations regarding the Coulomb interaction between electrons. In the
context of nanocooling a quantum dot by spin-polarized nonequilibrium currents,
a similar setup as suggested here has been investigated \cite{Jochen,Stadler}. 
There, the focus has been on the dynamical cooling and heating of the phonon
mode in presence of a {\em magnetic\/} molecular junction. 

In this work, we consider a nonmagnetic molecular bridge with weak
coupling to ferromagnetic leads. We include, besides the electronic degrees of
freedom, also a coupling of the electrons to a single mode of a molecular
vibration. Furthermore, we restrict our analysis to
the sequential tunneling regime, which we explore by means of a real-time
diagrammatic approach for the reduced density matrix of the molecular bridge.
After
 introducing the ferromagnetic Anderson-Holstein model in
Sec.~\ref{sec:ahmodel}, we discuss the necessary modifications in
the quantum kinetic equation due to the spin-polarized leads, as opposed to the
case of non-magnetic leads. Most importantly, we find an exchange magnetic field
which acts on the dot electrons and which is generated by the coupling of the
electrons to the vibrational mode. We discuss this effect for realistic 
experimental parameters in Sec.~\ref{sec:exfield}. We
calculate the relevant spin observables and discuss spin 
precession and accumulation also in terms of specific elements of the
reduced density matrix in Sec.~\ref{sec:spin}. Transport observables (such as
the charge current) for a 
nonequilibrium situation as well as for different polarization angles of the
leads are presented in Sec.~\ref{sec:transport}. We show that the
phonon-induced exchange magnetic field generates a current blockade which is
weaker as compared to the pure spin-valve effect. To
complete the physical picture, we also provide the results for the general case
when both the Coulomb and the electron-phonon coupling are present. We
underpin these numerical findings in terms of an attractive effective Anderson
model for the polarons. We conclude and summarize our findings in
Sec.~\ref{sec:conclusion}.

\section{Model and Diagrammatic Technique}
\label{sec:ahmodel}
\begin{figure}[t]
\centering
\includegraphics[width=\columnwidth]{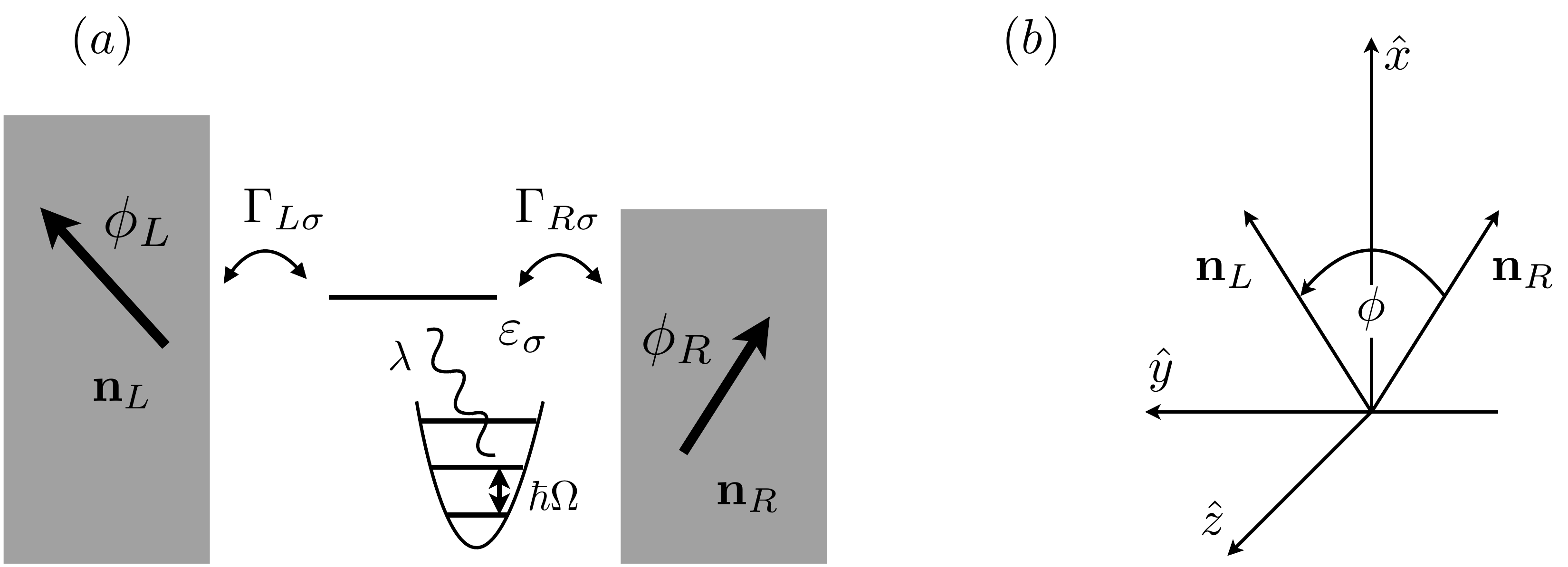}
\caption{\label{quant_ax} (a) Sketch of the Anderson-Holstein spin-valve
geometry.
The spinful electronic level with energy $\varepsilon_\sigma$ is coupled to a
single mechanical harmonic mode with frequency $\Omega$ with a coupling
constant $\lambda$. The electrons are able to tunnel from the spin-polarized
leads to the non-magnetic device due to finite tunneling couplings
$\Gamma_{L/R\sigma}$. The left
and right ferromagnets have different magnetizations, which are
characterized by an angle $\phi_\alpha,\alpha=L/R$. (b) The underlying
coordinate system where $\phi$ is the relative angle between the two
directions ${\bf n}_{L/R}$ of the left and the right magnetization.}
\end{figure}

In order to model the Anderson-Holstein spin-valve geometry, we use the
Hamiltonian  
\begin{equation}
\mathcal{H}=H_{mol}+H_T+H_{leads}. 
\end{equation}
The central region, see Fig.~\ref{quant_ax}(a), is denoted as '{\em dot\/}',
or, equivalently, as '{\em molecule\/}' in what follows and may be given as,
e.g., a single electronic level of a quantum dot or a
molecule. Its Hamiltonian reads ($\hbar =1$)  
\begin{equation}
H_{mol}=\sum_\sigma \epsilon_\sigma d_\sigma^\dag d_\sigma + Un_{\up}
n_{\down}+\Omega b^\dag b+ \lambda \hat{x} \left(d_{\up}^\dag
d_{\up} +d_{\down}^\dag d_{\down}\right).
\label{eq:ham_mol}
\end{equation}
The fermionic operators $d_\sigma(d^\dag_\sigma)$ annihilate (create) an
electron with spin $\sigma=\up/\down$ on the dot. Moreover, the electron number
operator on the dot is $n_{\sigma}=d^\dag_\sigma d_\sigma$. The single particle
energy is given by $\epsilon_\sigma$ and may be controlled by an applied gate
voltage
and/or an external magnetic field. In the absence of an additional external
magnetic field, we have that $\epsilon_{\up}=\epsilon_{\down}=\epsilon_0$.
Onsite Coulomb interaction is taken into account by the parameter $U$.
Furthermore, the vibrational mode is characterized by the bosonic operators
$b/b^\dag$, and its frequency is denoted by $\Omega$. The
(spin-independent) electron-phonon coupling strength is provided by $\lambda$.
The oscillator displacement operator is denoted by $\hat{x}$, and
the harmonic eigenstates are labeled by the quantum numbers $|m\rangle$. In the
absence of the coupling to the phonon, $\lambda=0$, the electronic eigenstates 
$\{|0\rangle, |\up\rangle, |\down\rangle, |d\rangle\}$ correspond to the empty,
the two singly and the doubly occupied state of the quantum dot.

Spin-dependent transport features emerge due to ferromagnetic leads. We allow
for different hybridizations for the up- and the down-spin electrons between
the leads and the dot and describe them by the Hamiltonian 
\begin{equation}
H_{leads}=\sum_{{\bf k}\alpha\tau}\left(\epsilon_{\bf
k\tau}-\mu_\alpha\right)c_{{\bf k}\alpha\tau}^\dag c_{{\bf k}\alpha\tau}\, .
\end{equation}
Electrons in lead $\alpha=L/R$ with energy
$\epsilon_{k\tau}$ are created (annihilated) by the fermionic operators
$c^\dag_{{\bf k}\alpha\tau} (c_{{\bf k}\alpha\tau})$, where $\tau=\pm$ labels the
majority/minority spin species. The quantization
axis of the left and right lead, respectively, is defined by the respective
magnetization directions ${\bf n}_L$ and ${\bf n}_R$. Both enclose an
angle $\phi_\alpha$ with the $x$-axis, see Fig.~\ref{quant_ax}(b). We denote by
$\mu_\alpha$ the electro-chemical potential of the lead $\alpha$. Ferromagnetism
is included by assuming different densities of states $\rho_{\tau=\pm}$ at the
Fermi energy for majority/minority spin electrons in the respective
lead. The polarization of the ferromagnetic lead $\alpha$ is defined by  
$p_\alpha=\frac{\rho_{\alpha+}-\rho_{\alpha-}}{\rho_{\alpha+}+\rho_{
\alpha-}}$. For nonmagnetic leads, $p_\alpha=0$, whereas $p_\alpha=1$
corresponds to a fully polarized lead hosting majority spins only. The dot and
the leads are tunnel coupled with an amplitude $t_\alpha$, leading to the
Hamiltonian \cite{Koenig3, Koenig2}
\begin{equation}
H_T=\sum_{{\bf k}\alpha\sigma\tau}\left\{t_\alpha c^\dag_{{\bf
k}\alpha\tau}U^{\alpha}_{\tau\sigma}d_{\sigma}+H.c\right\},
\end{equation}
where the unitary operator $U^{\alpha}_{\tau\sigma}$ describes the
rotation of the spin coordinate system of the on-dot electron to the common
reference frame. As in Ref.~\onlinecite{Koenig2, Koenig3}, we choose the
quantization axis for electron spins on the dot along the $z$-axis of the system
that is
spanned by ${\bf e}_x=({\bf n}_R+{\bf n_L})/|{\bf n}_R+{\bf n_L}|$, ${\bf
e}_y=({\bf n}_R-{\bf n_L})/|{\bf n}_R-{\bf n_L}|$, and ${\bf
e}_z={\bf e}_x\times {\bf e}_y$, see also Fig.~\ref{quant_ax}(b).  
In passing, we note that for spinors which are rotated by an angle $\theta$
about an axis parallel to the direction of the unit vector $\hat{{\bf n}}$,  the
following transformation holds
\begin{equation}
e^{i(\theta/2)(\hat{\bf n}\cdot {\bm \sigma})}=\cos(\theta/2)+i\hat{\bf
{n}}\cdot {\bm \sigma}\sin(\theta/2),
\end{equation}
where ${\bm \sigma}=(\sigma_x,\sigma_y,\sigma_z)$ is the vector of the Pauli
matrices.
Accordingly, the coordinate system is rotated by the opposite angles,
such that (see Fig.~\ref{quant_ax}(b)) the tunneling Hamiltonian assume the
form  
\begin{align}
H_T&=\sum_{{\bf k}\alpha}t_{\alpha}\left[c^\dag_{{\bf k}
\alpha+}\left(e^{i\phi_\alpha/2}d_{\up}+e^{-i\phi_\alpha/2}d_{\down}
\right)\right.\nonumber\\
& \left.c^\dag_{{\bf k}
\alpha-}\left(-e^{i\phi_\alpha/2}d_{\up}+e^{-i\phi_\alpha/2}d_{\down}
\right)\right]+H.c.
\end{align}
Here, we have used two rotations in series about the $z$- and $y$-axis,
respectively, such that 
$U_{\tau\sigma}=e^{i(\phi_\alpha/2)\sigma_z}e^{i(\pi/2) \sigma_y}$.
For a finite difference in the electrochemical potentials between the left and
right lead, $eV=\mu_L-\mu_R$, a net charge current will flow and the system
is in general in a nonequilibrium situation. For $p_\alpha\neq 0$, the
hybridization between the dot and the lead $\alpha$ becomes in general spin
dependent and we may introduce the tunneling coupling strength 
$\Gamma_{\alpha\pm}=2\pi|t_\alpha|^2\rho_{\alpha\pm}$. Hence, electronic states
with spin $\sigma=\up/\down$ acquire a finite line width
$\Gamma_\alpha=\sum_{\sigma=\pm}\Gamma_{\alpha\sigma}/2$ due to their
hybridization with the reservoirs.

Before we proceed with the calculation of the nonequilibrium density
matrix, we apply a polaron transformation, such that the electrons and the 
phonon decouple according to $H'=SHS^\dag$ \cite{Mahan,Karsten1}. 
Here, 
$S=e^{(\lambda/\Omega)(b^\dag-b)\sum_\sigma d^\dag_\sigma d_\sigma}$. By this, 
the eigenenergies of the bare electronic system are shifted by the polaron
energy, such that  $\epsilon_\sigma\to \epsilon_\sigma-\lambda^2/\Omega$. 
Successively, also the Coulomb repulsion strength gets shifted: $U\to
U-2\lambda^2/\Omega$. In passing, we note that typical terms of the tunneling
Hamiltonian transform under $S$ according to $d_\sigma\rightarrow d_\sigma
e^{(\lambda/\Omega) (b^\dag-b)}$ and, correspondingly 
for the creation operators, see Ref.~\onlinecite{Karsten1} for the
details. 

Since we are interested in the physical properties of the dot in the first
place, the lead and the bosonic degrees of freedom are traced out, thereby
yielding the reduced density matrix of the system. In order to proceed
analytically, we exploit the fact that the lead and phonon Hamiltonians are
quadratic in the fermionic and bosonic operators. By performing the trace, we
implicitly assume that both of these subsystems remain in thermal equilibrium,
even in the presence of a finite bias voltage. Otherwise, dynamic nonequilibrium
occupations of the phonon states and/or the lead electronic states have to be
explicitly taken into account as well before tracing over them. The reduced
density matrix of the system in general has six non-vanishing entries and
includes the coherences due to the coherent spin evolution driven by the
ferromagnetic leads. By following the steps described
in Refs.\ \onlinecite{Koenig1,Sothmann1}, we may obtain the kinetic equation for
the elements of the reduced density matrix  
\begin{eqnarray}
\frac{d}{dt} \rho^{\chi_1}_{\chi_2}(t) =-i \sum_\chi \left(
h_{\chi_1\chi}\rho^{\chi}_{\chi_2} -
	h_{\chi \chi_2} \rho^{\chi_1}_{\chi} \right) (t) +
	\nonumber \\
	\sum_{\chi_1' \chi_2'} \int_{-\infty}^t dt' \Sigma_{\chi_2
\chi_2'}^{\chi_1 \chi_1'} (t,t') \rho_{\chi_1'}^{\chi_2'} (t') 
	\, ,   
 \label{rhodot}
\end{eqnarray}
with the irreducible self energy kernels
$\Sigma_{\chi_2,\chi_2'}^{\chi_1,\chi_1'}(t,t')$. They include transitions
of the system between even and odd parity states that are induced by the
tunneling Hamiltonian. The internal coherences  are covered by the first term
and are generated by the molecular Hamiltonian $H_{mol}$. They are given by the
Hamiltonian matrix elements $h_{\chi\chi'}$ of $H_{mol}$, with
$\chi,\chi'\in\left[\{|0\rangle,|\up\rangle,|\down\rangle,|d\rangle\}
\otimes |m\rangle\right]$, of the dot-plus-phonon-system in the absence of the
coupling to the leads. We note that for a spin-independent electron-phonon
interaction $\lambda$, a contribution to
zeroth order in the hybridization $\Gamma_\alpha$ arises, which is diagonal in
the basis of the phonon. 

To determine the irreducible self energy kernels, we employ the real-time
diagrammatic technique
\cite{Koenig1,Sothmann1} which gives a systematic expansion of the kinetic
equation in orders of $\Gamma_\alpha$.  The first non-vanishing order includes
all terms which 
change the number of electrons on the dot by one. For practical purpose, we
shall include the maximal number $n$ of states in the bosonic Hilbert space and
verify convergence with respect to increasing $n$. Then,  all $4n$ diagonal
elements of the density matrix, are in general nonzero and represent the
occupations of the respective polaron states of the dot. In addition, the
overlap between the singly occupied states with spin-up and -down is
finite in the presence of ferromagnetic leads\cite{Koenig2,Koenig3}. In
the presence of the vibrational degree of freedom, the appearing Fermi functions
are dressed by the Franck-Condon matrix elements, see
Refs.~\onlinecite{Koenig1,Sothmann1}. They simplify if we assume the
vibrational mode to be always at thermal equilibrium
\cite{Mahan,Leijnse,Koch,Karsten1,Karsten2}. Then, we find 
\begin{equation}
F^{\pm}_\alpha(\Omega,\epsilon)=\int_{-\infty}^\infty
\frac{d\omega}{2\pi}f^\pm_\alpha(\omega+\epsilon)C_{\Omega}^\pm(\omega),
\end{equation}
where $C_{\Omega}^+(\omega)=2\pi\sum_{n=-\infty}^\infty
P_n(\lambda/\Omega)\delta(\omega-n\Omega)$ (see Ref.\ \onlinecite{Karsten2}) is
the Fourier
transform of the correlator 
\begin{equation}
C_{\Omega}^+(t)=\left\langle e^{-(\lambda/\Omega)\left(b^\dag e^{i\Omega
t}-be^{-i\Omega t}\right)}e^{-(\lambda/\Omega)(b^\dag-b)}\right\rangle.
\end{equation}
Moreover, we have the detailed balance relation
$C^+(\omega)=C^-(\omega)e^{\beta\omega}$.
For completeness, we comment that the $P_n(x)$ have an expansion in terms of
modified Bessel functions \cite{Mahan}. For practical reasons, infinite
sums are truncated
once convergence is achieved for the observables of interest. Typically
$n=20$ is sufficient for large $\lambda/\Gamma\simeq 10$ as studied in the
following.

Charge current and spin expectation values are the interesting
observables, which are also accessible in experimental setups. In the following,
we
focus on the symmetric case, $\Gamma_{L}=\Gamma_{R}$, and
$\mu_{L}=-\mu_{R}=eV/2$ as well as $p_L=p_R=p$ and
$\phi_L=-\phi_R=\phi/2$ (a generalization is straightforward). We present
results for the differential
conductance $dI/dV$ below, derived from the symmetrized tunneling current
$I=(I_L-I_R)/2$, with 
\begin{equation}
I_\alpha(t)=-e\frac{d}{dt}\sum_{{\bf k}\alpha\sigma}\langle c^\dag_{{\bf
k}\alpha\sigma}c_{{\bf k}\alpha\sigma}\rangle.
\label{Isymm}
\end{equation}
We note that the current operator has a similar structure as the tunneling
Hamiltonian. Within the diagrammatic formalism, we have to replace a
tunneling vertex by a current vertex and use the corresponding transition rates
$\Sigma_{\,\chi_2,\chi_2'}^{I \chi_1,\chi_1'}(t,t')$, for details see
Ref.\ \onlinecite{Sothmann1}. Information about the spin state of the system
is obtained from the reduced density matrix $\rho_\chi^{\chi'}$, which is 
related to the vector of the average spin ${\bm
\sigma}=(\sigma_x,\sigma_y,\sigma_z)$. 
We then have for the averages of the spin projections 
\begin{align}
\langle
\sigma_{x}\rangle=\frac{\rho_{\downarrow}^{\uparrow}+\rho_{\uparrow}^{
\downarrow}}{2},
\langle
\sigma_{y}\rangle=i\frac{\rho_{\downarrow}^{\uparrow}-\rho_{\uparrow}^{
\downarrow}}{2},
\langle
\sigma_{z}\rangle=\frac{\rho_{\uparrow}^{\uparrow}-\rho_{\downarrow}^{
\downarrow}}{2}.
\end{align} 

\section{Exchange magnetic field}
\label{sec:exfield}
\begin{figure}[t]
\begin{center}
\includegraphics[width=1.2\columnwidth]{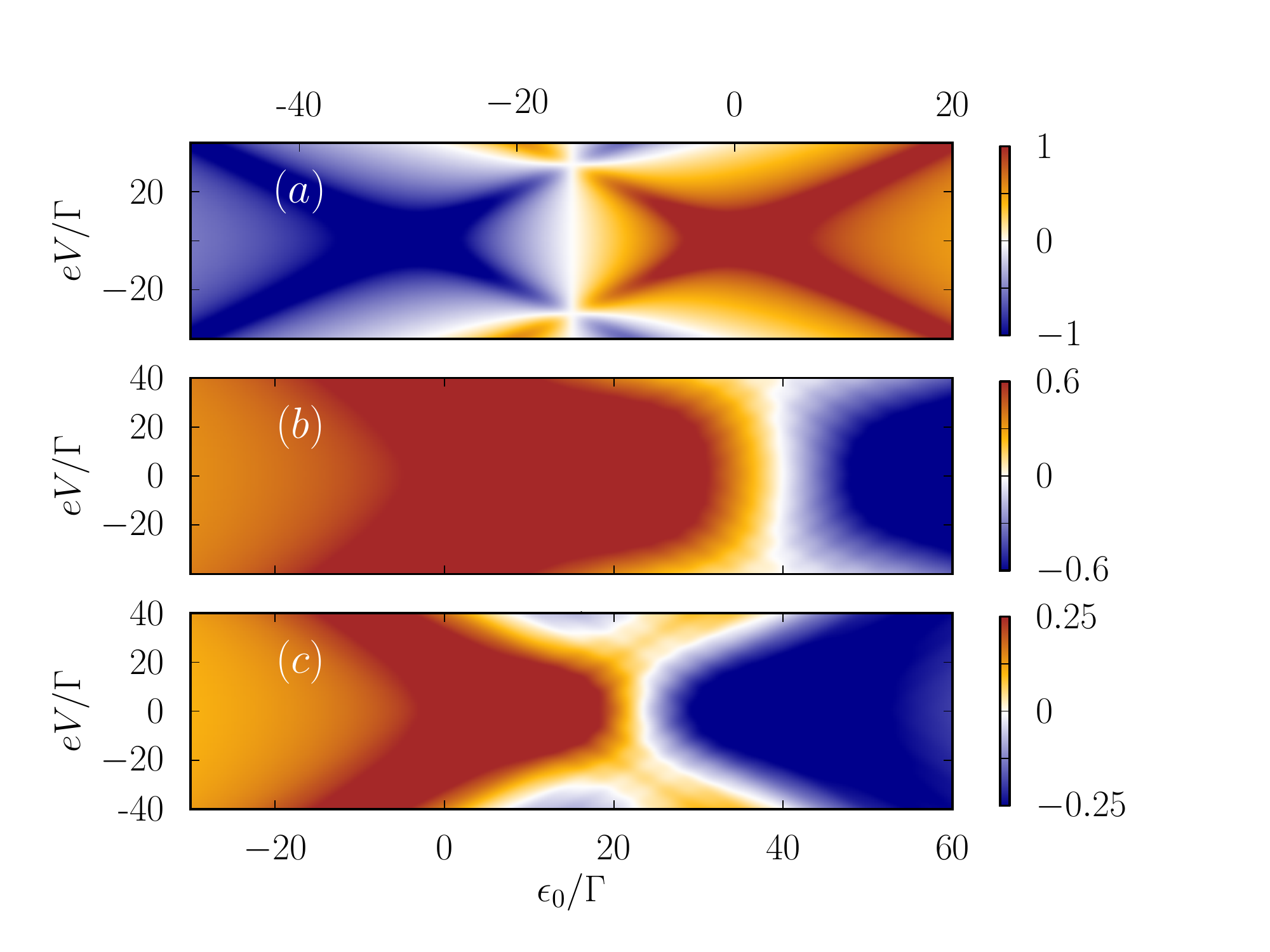}
\end{center}
\caption{\label{exchange} Effective exchange magnetic field in units of
$\Gamma$ induced by the
electron-phonon and Coulomb interaction. 
The color scale shows the sum $B_{\rm ex}=B_{L}+B_{R}$. Panel (a) shows the
result for pure Coulomb interaction $U=30k_BT$ without electron-phonon coupling
$\lambda=0$. In (b), the electron-phonon coupling is set to 
$\lambda=10 k_BT,\Omega=5k_BT$, while $U=0$. The finite $\lambda$ broadens the resonances, 
shifts it to larger values of the gate voltage and reverses
its direction as compared to panel (a). In (c), both interactions are present.
The remaining parameters are $k_BT=\Gamma$, $\phi=\pi/2$ and $p=0.9$.}
\end{figure}
Ferromagnetic leads induce a coherent spin dynamics on the dot although the
quantum dot itself is nonmagnetic. In lowest
order in the lead-dot tunneling coupling, a contribution to an effective
 exchange magnetic field is generated from each of the two leads. It points in
the respective direction of the lead magnetization. We follow the
derivation in Refs.\ \onlinecite{Koenig2, Koenig3} and explore the fact that 
the charge as well as the spin degrees of freedom obey rate equations. The
dynamics of the spin can be formulated in terms of a time evolution equation
for the spin vector ${\bm \sigma}(t)$ which
describes the dynamics of the spin of the confined electron. This vector
equation can be decomposed into three parts according to 
\begin{equation}
\frac{d{\bm \sigma}}{dt}=\left(\frac{d{\bm \sigma}}{dt}\right)_\text{acc}+\left(\frac{d{\bm \sigma}}{dt}\right)_\text{rel}+\left(\frac{d{\bm \sigma}}{dt}\right)_\text{rot},
\label{spineq}
\end{equation}
where accumulation, relaxation and rotation of the spin are identified by the three terms. 
Correspondingly, the dynamical equation of motion for the charge degrees of
freedom can be decomposed in a similar manner. As opposed to
Refs.~\onlinecite{Koenig2,Koenig3}, all energies or differences 
of energies between singly and doubly occupied states have to be replaced by the
corresponding energies of the respective polaron states, see
Sec.~\ref{sec:ahmodel}. A closer look to the third 
term in  Eq.~\eqref{spineq} yields the representation 
\begin{equation}
\left(\frac{d{\bm \sigma}}{dt}\right)_\text{rot}={\bm \sigma}\times \left({\bf
B}_L+\bf{B}_R\right)
\end{equation}
in terms of an effective exchange magnetic field with the contributions 
 ${\bf B}_\alpha$\cite{Koenig3}. It is well known that for standard
spin-valves without a vibrational mode, the Coulomb interaction $U$ between
electrons
 can generate a finite exchange magnetic field (see Eq.~\eqref{Bex} below for
$\lambda=0$). In presence of the electron-phonon coupling, the vibrational
coupling generates an effective attractive Coulomb interaction. In
turn, the resulting exchange magnetic field depends on $\lambda$ as well as on
$U$. Overall, we find the analytic expression for the exchange magnetic field 
\begin{equation}
 {\bf
B}_\alpha=-\frac{p_\alpha\Gamma_\alpha}{\pi}\left[
\Lambda_\alpha\left(\epsilon-\frac{\lambda^2}{\Omega}
\right)-\Lambda_\alpha\left(\epsilon+U-\frac{3\lambda^2}{\Omega}\right)\right]{
\bf n}_\alpha,
 \label{Bex}
\end{equation}
which acts from contact $\alpha$ on the spin of the electron confined on the
dot. Here,
$\Lambda_\alpha(x)=\text{Re}\,
\Psi\left[1/2+i\beta(x-\mu_\alpha)/(2\pi)\right]$,
with the Digamma function $\Psi$ \cite{Gradshteyn}. It is important to realize
that even in the absence of the Coulomb interaction, i.e., for $U=0$, a finite
exchange field ${\bf B}_\alpha$ in the direction ${\bf n}_\alpha$ of the
magnetization of the lead $\alpha$ exists when an 
electron-phonon coupling is present, $\lambda\neq 0$. Consequently, the spin of
a confined electron, which is subject to electron-phonon coupling, will precess
in the effective exchange magnetic field generated when $\lambda \ne 0$. 
Eq.~\eqref{Bex} includes previous results in the case of $\lambda=0$.
Furthermore, it is interesting to realize that for Coulomb interaction
strengths $U<2\lambda^2/\Omega$, the exchange magnetic 
field points into the direction opposite to that of the lead magnetization.
Moreover, it vanishes exactly when $U=2\lambda^2/\Omega$. The exchange magnetic
field is shown in Fig.~\ref{exchange} as a function of the bias and the gate
voltages in a color scale plot in units of $\Gamma$ for three different cases.
In
Fig.~\ref{exchange}(a), only pure Coulomb interaction is present while the
electron-phonon coupling is switched off ($U=30 k_BT,\lambda=0$). In turn, in 
Fig.~\ref{exchange} (b), only pure electron-phonon interaction is present while
the Coulomb term is absent ($U=0,\lambda=10 k_BT$). Here, we have 
chosen the frequency of the oscillator as $\Omega=5k_BT$. Finally, in
Fig.~\ref{exchange} (c), both interactions are present
($U=30 k_BT,\lambda=10 k_BT$). In the presence of the 
Coulomb interaction alone and  for a fixed bias voltage, the exchange field
shows sharp resonances. In between these resonances, the field decreases, goes
through zero, reverses its sign, and increases again to the adjacent resonance
in Fig.~\ref{exchange}(a). For pure
electron-phonon coupling and in absence of a Coulomb interaction $U=0$ (see
Fig.~\ref{exchange}(b)), the analogy to an effective attractive Coulomb
interaction appears in two respects: First, there is a finite exchange
field induced by the phonon on the dot. Second, as opposed to (a) it has a
reversed sign. As in (a), we find two resonances, which are shifted to higher
values of $\epsilon_0$ here. This is a general feature which is due to the
opposite sign of the effective Coulomb interaction constant. In panel (c), 
the parameters are chosen such that $\lambda^2/\Omega \approx U$. The exchange
magnetic field is still visible, although being smaller in
amplitude as opposed to (a). The width of the resonance is
broader than in (a) which reflects the dominance of $\lambda$. For those
parameters, we have that $\lambda^2/\Omega>U/2$ and the tendency is as in
Fig.~\ref{exchange}(b). In contrast, the choice $U/2>\lambda^2/\Omega$ would
result in a reversed direction of the effective exchange field again. In what
follows, we discuss how the results are modified in comparison to the case 
$U\neq 0, \lambda=0$ by a finite electron-phonon coupling.

\section{Spin precession and spin accumulation}
\label{sec:spin}
\begin{figure}[t]
\begin{center}
\includegraphics[width=\columnwidth]{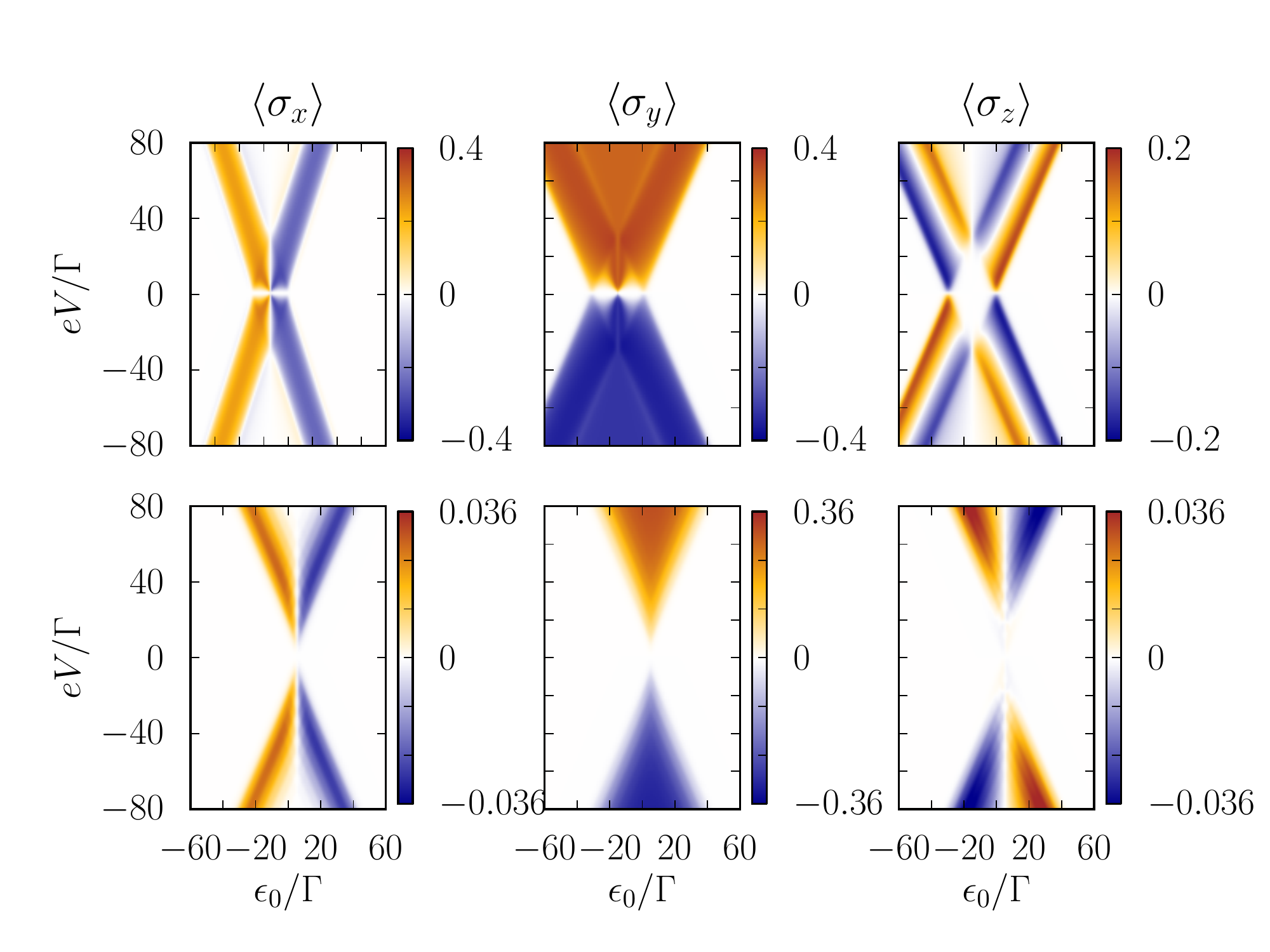} 
\end{center}
\caption{\label{sexp}Interaction-induced spin accumulation illustrated by 
the spin expectation values $\langle
\sigma_x\rangle$ (left column),  $\langle \sigma_y\rangle$ (middle column) and $\langle
\sigma_z\rangle$ (right column) for varying bias and
gate voltages and for $U=30 k_BT$. The upper row shows the results for pure
Coulomb interaction without electron-phonon coupling $\lambda=0$. The
lower row refers to the case with an additional electron-phonon
interaction $\lambda=10 k_BT, \Omega=5k_BT$. The remaining parameters are
as in Fig.~\ref{exchange}.}
\end{figure}
The immediate consequence of the interaction-induced exchange magnetic field is
a spin accumulation and spin precession of single electron spins on the dot.
Whenever the dot is doubly occupied, these effects are expected to be less
pronounced, since the spin singlet state is not sensitive to inhomogeneous
magnetic fields. Spin measurements have been performed in order to
determine the strength of the exchange magnetic field, e.g., in carbon nanotubes
contacted by ferromagnetic leads \cite{Hauptmann}. In the following, we address 
 the spin expectation values along the three spatial directions with the
coordinate system being defined as indicated in Fig.~\ref{quant_ax}(b). We show 
results for the angle $\phi=\pi/2$ between the left and right magnetization
directions. This corresponds to the configuration ${\bf n}_L\perp {\bf n}_R$.
From Refs.~\onlinecite{Koenig2,Koenig3} it is known that for small bias
voltages, a finite spin accumulation occurs along the direction  ${\bf
n}_L-{\bf n}_R={\bf e}_y$. It also occurs in the absence of the
exchange field. It is important to realize that a finite $\langle \sigma_x\rangle\neq
0$ and/or $\langle \sigma_z\rangle\neq 0$ is a clear signature of a finite
exchange field generated by Coulomb or electron-phonon interactions. 

The induced spin accumulation is illustrated in Fig.~\ref{sexp} by the spin
expectation values $\langle \sigma_x\rangle,\langle \sigma_y\rangle$ and $\langle
\sigma_z\rangle$ for varying bias and gate voltages. The upper
row shows the results for the case of pure Coulomb interaction, 
$U=30k_BT$ and $\lambda=0$. For $\phi=\pi/2$, spin is accumulated in
the direction ${\bf n}_L-{\bf n}_R$ due to the conservation of the total
angular momentum. This maximal spin-valve effect competes with the
spin precession around the induced exchange field which points in the direction
${\bf n}_L+{\bf n}_R$. The latter induces finite $x-$ and $z$-components of the
spin. Since two addition energies are present, also two distinct resonance
lines appear in the upper left panel. The same substructure also emerges in the
upper middle panel. In this case, the transitions on the dot between the states 
$|\up\rangle$ and $|\down\rangle$ and the empty or doubly occupied state are
non-degenerate due to the finite spin accumulation. 
In the lower row of Fig.~\ref{sexp}, we depict the results for
$U=10k_BT$ and $\lambda=10k_BT$ which illustrate the
modification of the spin accumulation and the spin precession due to a finite
electron-phonon coupling $\lambda$. They are in accordance with 
Fig.~\ref{exchange}(c), where a finite exchange 
field is shown.  Although the spin expectation values are about one
order of magnitude smaller than the exchange field, the qualitative physical
behavior is the same. The electrons couple  to the phonon mode differently
depending on their spin due to the lead polarization. By this, an effective
spin-phonon coupling emerges. On the other hand, the electron-phonon
coupling compensates the repulsive Coulomb interaction and allows the dot to
be occupied by two electrons. 

\subsection{Angular dependence}
\begin{figure}[t]
\begin{center}
\includegraphics[width=\columnwidth]{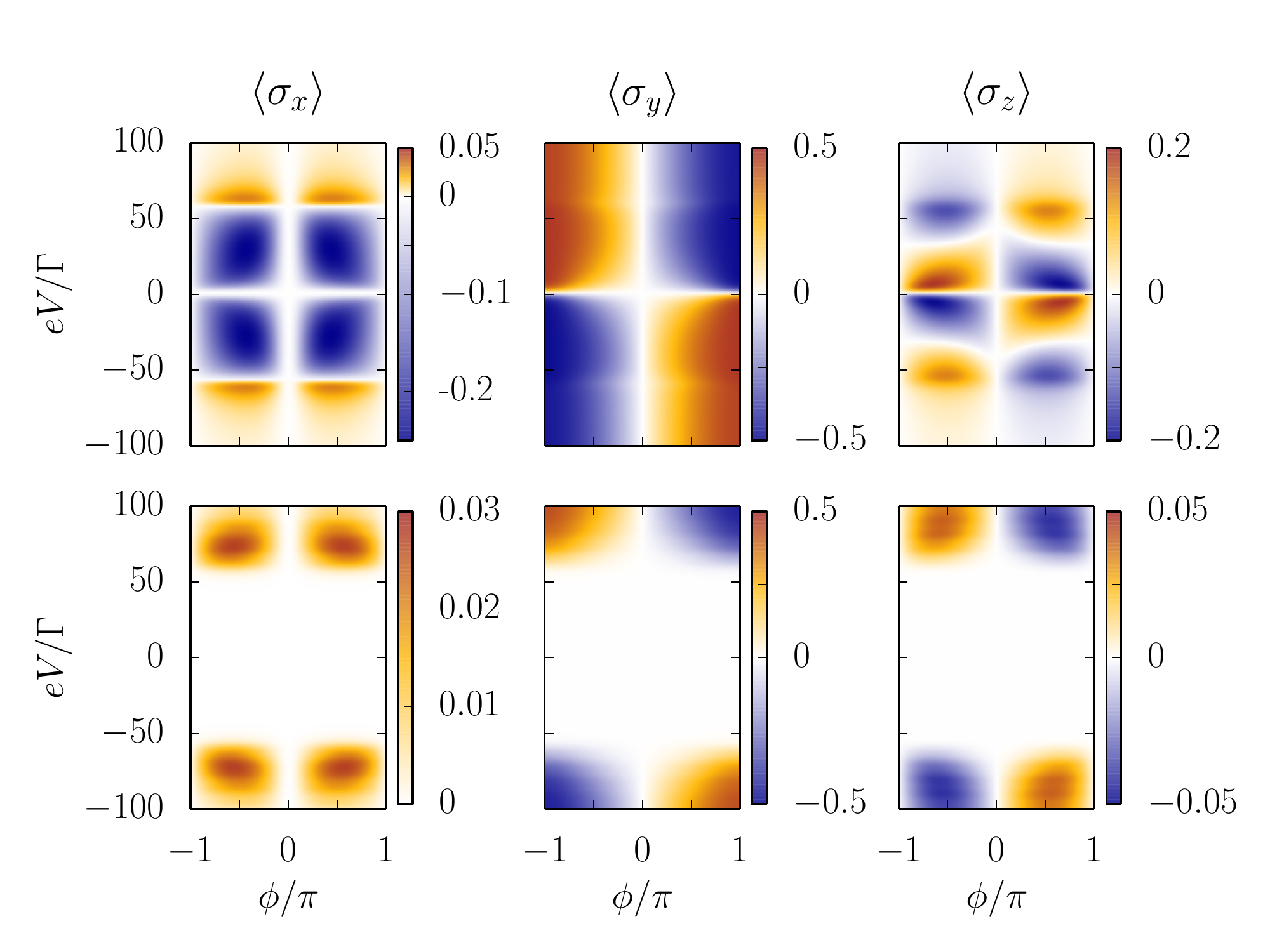} 
\end{center}
\caption{\label{sexp2} Same as in Fig.~\ref{sexp}, but as a function of the
enclosed angle $\phi$ for a single-particle energy $\epsilon_0=30k_BT$ and for
$U=30 k_BT$. The upper row shows the results for $\lambda=0$, while the lower
row shows the case with an additional electron-phonon interaction $\lambda=10
k_BT, \Omega=5k_BT$. }
\end{figure}
The choice of two independent magnetizations in the left and right lead,
characterized by the relative angle $\phi$, generates two non-collinear
magnetic fields for the spin of an electron on the dot. When the dot is
occupied by a single electron, the spin state is sensitive to these fields.
Then, also the dependence of the spin accumulation on $\phi$ is relevant. In
Fig.~\ref{sexp2}, we show the spin expectation values as a function of the
bias voltage and the enclosed angle $\phi$. We have fixed the gate voltage
to $\epsilon_0=30k_BT$ and the remaining parameters are as in Fig.~\ref{sexp}. 
Again, the three upper panels refer to pure repulsive Coulomb interaction with 
 $U>0$. The device is largely singly occupied and we find a clear
precession of the spin with finite $x$- and
$z$-components of the spin operator. In contrast to the spin accumulation, 
the spin precession vanishes for the collinear configurations, when $\phi=0, \pm
\pi$. As the spin-dependent effects are absent for zero bias voltage, we 
find that increasing the bias voltage favors a specific spin direction, either
up or down. This is accompanied by the precession of the spin with the same or
the opposite components of the spin vector. This is depicted in the left and
right panels of the upper row in Fig.~\ref{sexp2}. The results for finite
$\lambda$ are shown in the lower row. The spin accumulation is sensitive to
the change of the angle as well, similar as the behavior discussed above.
However, the characteristics of the precession is different, since it is
determined by the value of the exchange magnetic field. We note that 
always a positive $x$-component of the spin emerges, whereas the $z$-component
may reverse its sign, when either the angle or the bias voltage are changed from
positive to negative values. Since the Franck-Condon blockade is well developed
 in the central areas of the panels, spins of any kind are blocked to enter the
dot in this regime. Hence, neither spin accumulation nor spin precession occurs
here.

\section{Transport spectroscopy}
\label{sec:transport}
\begin{figure}[t]
\begin{center}
\includegraphics[width=\columnwidth]{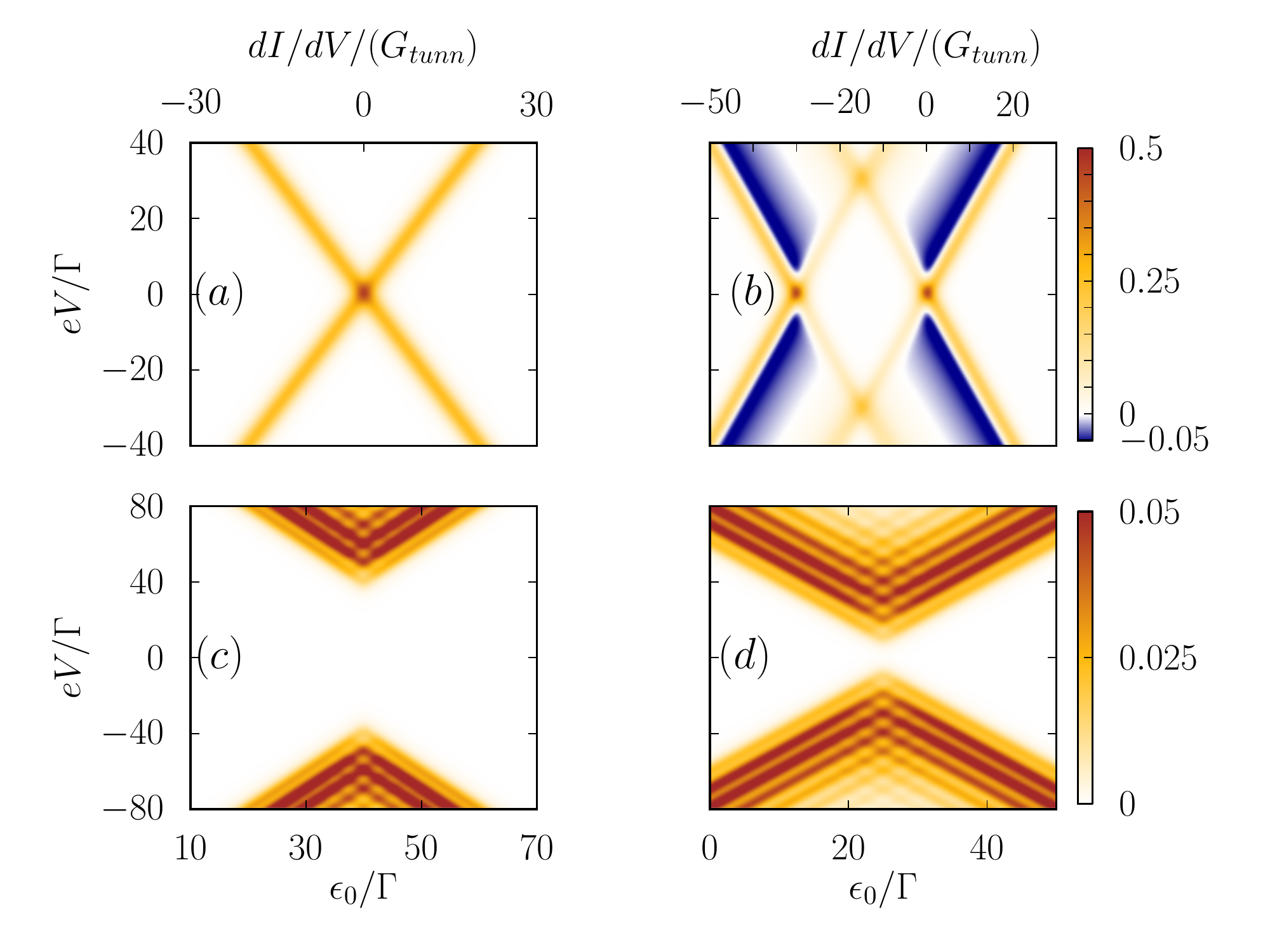} 
\end{center}
\caption{\label{dIdVand}Differential conductance of the Anderson-Holstein
spin-valve as a function of the bias and the gate voltage in units of
$G_{\rm tunn}=(e^2/h)\Gamma/(2\pi k_BT)$.  In (a), the noninteracting case is
shown: $U=\lambda=0$, while (b) refers to $U=30k_BT,\lambda=0$. In panel (c),
we set $U=0,\lambda=10k_BT$, and in (d), we
have chosen $U=30k_BT,\lambda=10k_BT, \Omega=5k_BT$. All calculations have been performed 
with $\Omega=5 k_BT$, $p=0.9$ and $\phi=\pi/2$.}
\end{figure}

\begin{figure}[t]
\begin{center}
\includegraphics[width=\columnwidth]{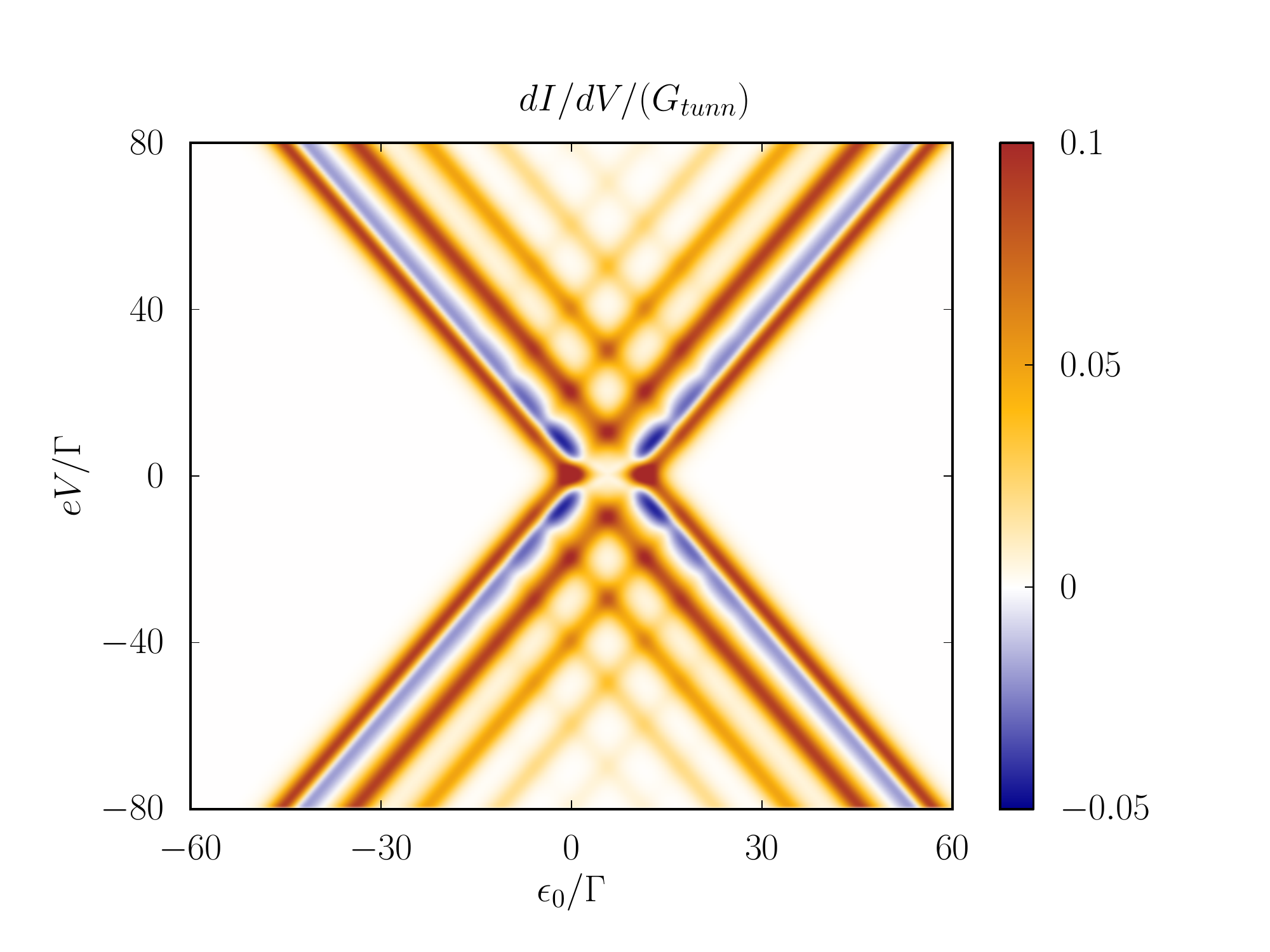} 
\end{center}
\caption{\label{ndIdVph}Differential conductance in the crossover regime (see
text) for $U=30k_BT,
\lambda=\Omega=10k_BT$. The remaining parameters are as in
Fig.~\ref{dIdVand}.}
\end{figure}
Measurements of the current give precise insight into the spectrum of the
interacting quantum 
dot. In this section, we present results for the differential conductance
$dI/dV$, with the current defined in Eq.~\eqref{Isymm}. The differential
conductance as function of $eV$ and $\epsilon_0$ is shown in
Fig.~\ref{dIdVand} (a) for $U=0,\lambda=0, \Gamma=k_{B}T$, and $\phi=\pi/2$. A
single conductance line is visible since the system possesses only one addition
energy
$E_{add}=\epsilon_{|d\rangle}-\epsilon_{|\uparrow/\downarrow\rangle}
=\epsilon_{|\uparrow/\downarrow\rangle}-\epsilon_{|0\rangle}=\epsilon_0$.
The finite conductance is accompanied by the inset of the spin accumulation on
the dot, see the discussion related to Fig.~\ref{sexp}. In Fig.~\ref{dIdVand}
(b), we depict the result for $U=30 k_BT,\lambda=0$. Without the ferromagnetic
leads, two addition energies exist in this case, $\epsilon_0$ and
$\epsilon_0+U$. Then, for $\epsilon_0\leq 10\Gamma$, the
dot is in the doubly occupied state, whereas for increasing gate voltages, the
singly occupied state is favored. This gives rise to the diamond in the center
region of Fig.~\ref{dIdVand} (b). For large gate voltages,
$\epsilon_0\gtrsim30\Gamma$, the dot is empty again. The line widths are
determined by the temperature, and we find pronounced satellite lines with
negative values. This negative differential conductance originates from a
partial blocking of spins, a competition between spin accumulation and
interaction-induced spin precession. Spin accumulation hinders electrons, which
do not have the right spin projection, to enter the drain, whereas spin 
precession counteracts to lift the blockade. The specific positions
of the lines are obtained from the shape of the induced exchange field, see the 
discussion in Sec.~\ref{sec:exfield}. With the lead magnetizations chosen such
that $\phi=\pi/2$, a maximal spin-valve effect is present. The charge degeneracy
points are located at $\pm\epsilon_0=-U/2$, where a maximal current flows. In
panel (c) of Fig.~\ref{dIdVand}, we show a typical Franck-Condon differential
conductance. There is a large range of bias and  gate voltages $(|eV|\leq
40k_BT, 30k_BT\leq|\epsilon_0|\leq 50k_BT)$, where transport is largely blocked.
Once the bias voltage provides enough energy, which is of the order of the
polaron energy $\lambda^2/\Omega$, electrons are able to tunnel through the dot
and a current can flow. We see equidistant lines in the spectrum that are 
associated with the energy differences between neighboring states. The regular
appearance could be explained in terms of
Franck-Condon parabolas that allow or forbid transitions from one vibrational
state to another, when at the same time the number of electrons is changed
\cite{Koch,Karsten1,Karsten2}. Due to the polaronic energy shift
$\epsilon_0\to\epsilon_0+\lambda^2/\Omega$, charge state degeneracies are
shifted differently as compared to the situation shown in
Fig.~\ref{dIdVand}(b). In particular, the shift occurs to
positive values of $\epsilon_0$. The effective attractive Coulomb
interaction tends to put the dot in the doubly occupied state. Clearly,
 features of a negative differential conductance as in Fig.~\ref{dIdVand} (b)
are absent then. In Fig.~\ref{dIdVand} (d), we
depict results for the scenario that both interactions are finite, i.e., 
$U=30k_BT$ and $\lambda=10k_BT$. For this particular choice
of parameters, the polaron energy approximately equals the Coulomb
repulsion and we obtain a competition between the effects of the regimes
described in panel (b) and (c). The parameters are $\lambda=\Omega=10k_BT,
U=30k_BT$. Decreasing the polaron energy further results in a crossover to the
behavior shown in Fig.~\ref{dIdVand}(b). There, the negative differential
conductance is fully developed, whereas the Franck-Condon steps vanish
successively. In Fig.~\ref{ndIdVph}, we depict the results for the crossover
regime, where we
find a superposition of modulated negative differential conductance lines
(blue) and Franck-Condon steps in the $dI/dV$-curve. 

\subsection{Angular dependence} 
\begin{figure}[t]
\begin{center}
\includegraphics[width=\columnwidth]{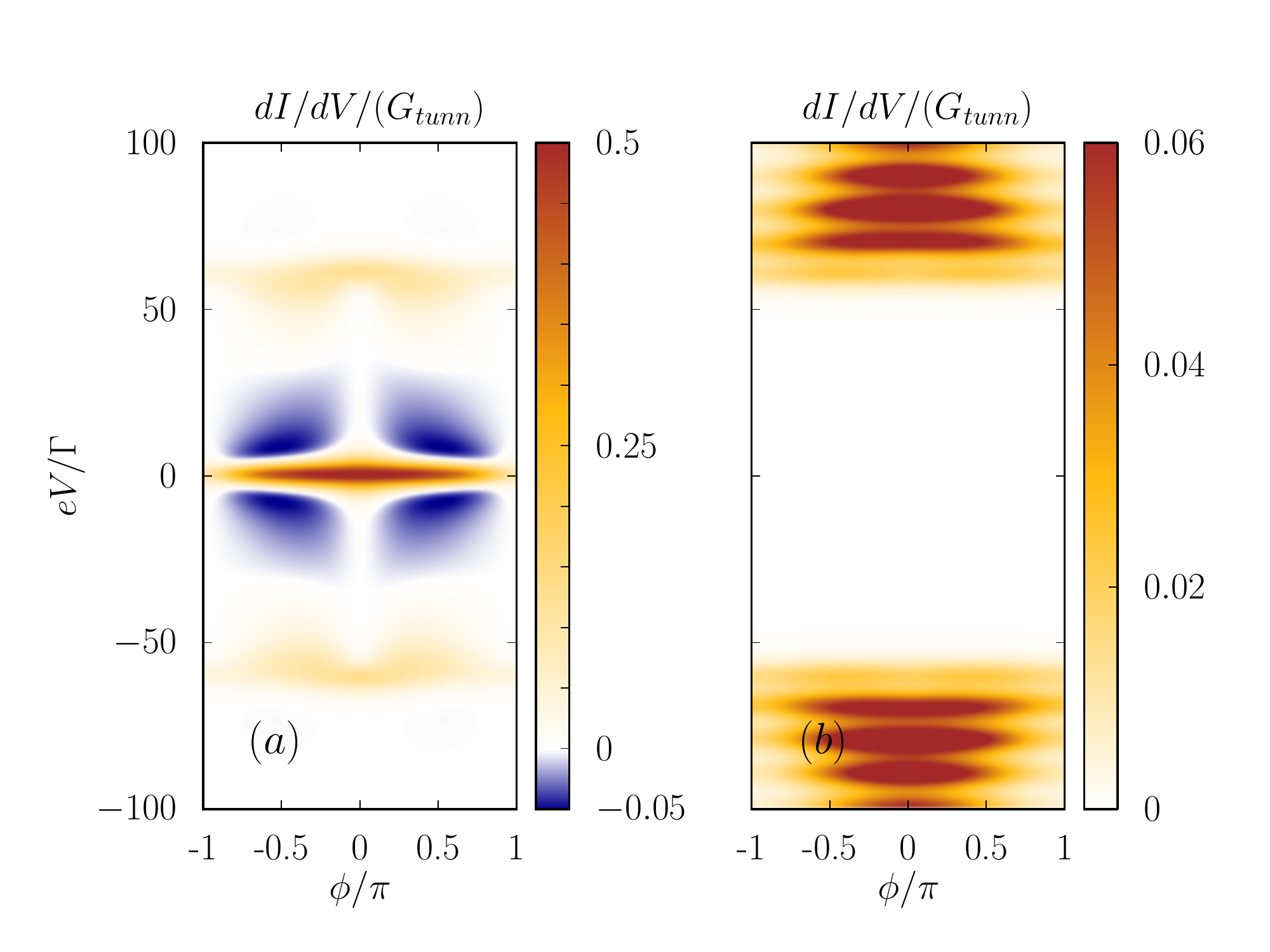}
\end{center}
\caption{\label{dIdVang}Angular dependence of the differential conductance
for $U=30k_BT$ without electron-phonon coupling $\lambda=0$ in (a), and
with $\lambda=10k_BT,\Omega=5k_BT$ in (b). The remaining parameters are as in
Fig.~\ref{dIdVand}.}
\end{figure}
The transport spectrum $dI/dV$ as a function of the noncollinearity of the lead
magnetizations $\phi$ is shown for a fixed gate voltage $\epsilon_0=30 k_BT$ in
Fig.~\ref{dIdVang}. For a vanishing electron-phonon coupling, $dI/dV(\phi,eV)$
for $\lambda=0$ is shown in Fig.~\ref{dIdVang}(a). The dot is essentially 
unoccupied. The symmetry of the figure is due to the symmetry of the underlying
setup, and we recall that we calculate the symmetrized current. For small bias
voltages, a positive conductance peak appears for varying $\phi$. Along
$\phi$, we recover the known suppression of the conductance away from the
maximum according to $\sin^2\phi$, see Refs.\ \onlinecite{Koenig2,Koenig3}. With
increasing bias
voltage, the dot is more populated. Depending on the
noncollinearity, spin is accumulated on the dot and precesses. Transport is then
reduced, and hence, a clear negative differential conductance feature occurs.
For even larger bias voltages, transport is independent of $\phi$. In 
Fig.~\ref{dIdVang} (b), we show the results for $U=30k_BT$ and
$\lambda=10k_BT$. Again, conductance lines are symmetric with respect to 
$\phi=0, eV=0$, and the strong Franck-Condon blockade is independent of the
enclosed angle. Since the electron-phonon coupling favors a doubly occupied dot
state, a spin singlet state is formed. This singlet is independent of any
magnetic field and thus is not influenced by changes of the angle of the lead
magnetizations. As a function of $\phi$, the respective lines show a maximum for
$\phi=0$, and the reduction follows again a $\sim\sin^2\phi$ functional form. 

\section{Nonequilibrium phonon mode}
\begin{figure}[t]
\begin{center}
\includegraphics[width=0.7\columnwidth]{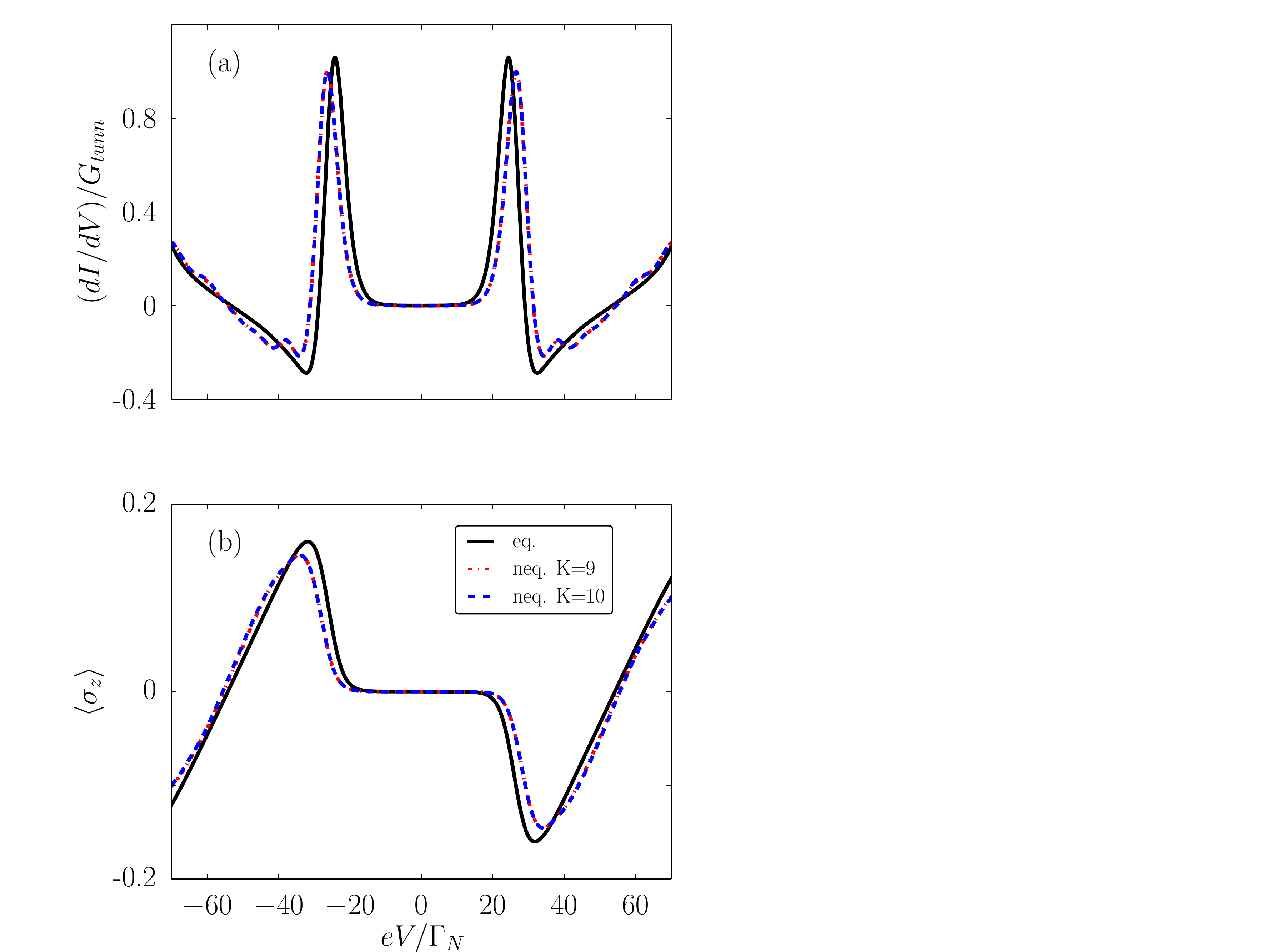}
\end{center}
\caption{\label{fig:nephonon}(a) Comparison of the differential conductance for 
equilibrated and non-equilibrated phonon modes. Within a voltage range 
$|eV|\leq 20\Gamma$ we do not find significant deviations from the 
calculation where the phonon is assumed to be in equilibrium. 
(b) Same as in (a) but for the nontrivial spin projection along the $z$-axis. 
We note that $\langle \sigma_z\rangle\neq 0$ reflects the finite 
exchange magnetic field from the leads. Other parameters are choses as 
$\lambda=2.5\Gamma, k_BT=\Gamma, \phi=\pi/2,
\Omega=5\Gamma,\epsilon=-40\Gamma$.}
\end{figure}
In the discussion above, we have assumed that the phonon mode is always kept at
thermal equilibrium. In this section, we address the situation when this no
longer holds and the phonon mode is treated fully nonadiabatically. Starting
from the Hamiltonian in Eq.~\eqref{eq:ham_mol}, we have to truncate the phonon
Hilbert space at a certain number $K$ of energy eigenstates. When calculating
the kernels within the real-time diagrammatic approach, we explicitly keep the
dependence on these basis states. Clearly, the truncation of phonon
absorption and emission processes becomes increasingly cumbersome for large 
electron-phonon coupling. Especially in the context of cooling or heating 
the environment, see Ref.~\onlinecite{Jochen}, the full density matrix of the
electron and phonon degrees of freedom is of importance in order to monitor the
effective temperature which refers to the mean energy contained in the
vibrational mode. In order to include nonequilibrium effects in the phonon
subspace, we trace over the bosonic degrees 
of freedom only when we compute electronic observables, such as the current or
the spin expectation values. Otherwise, we keep the $(4K)^2$ 
elements of the density matrix of the dot. We show in Fig.~\ref{fig:nephonon}(a)
the differential conductance as a function of the bias voltage for
$\lambda=2.5\Gamma,\Omega=5\Gamma, k_BT=\Gamma, U=30\Gamma$, and
$\epsilon=-40\Gamma$. The solid line refers to the results with the assumption
of a thermally equilibrated phonon, whereas the dashed 
lines mark the results for different numbers of phonon states that are taken
into account. The red (blue) line is for $K=9 (K=10)$ phonon states. 
We find a good agreement for small to intermediate bias voltages
$0\leq |eV|\leq 30\Gamma$. Small side peaks develop at larger bias
voltages, which are associated to the influence of the nonequilibrium
phonon distribution. However, the key features, such as the appearance of the
negative differential conductance and the Lorentzian peaks at 
$|eV|=U/2+\lambda^2/\Omega$ are reproduced  by the 
equilibrium phonon distribution as well. The height of the peaks differ to
some extent between the equilibrium and the nonequilibrium phonon model. For
larger values of $\lambda>3\Gamma$, it is difficult to obtain
numerically converged results and the calculation in the
Franck-Condon regime become increasingly cumbersome. In
Fig.~\ref{fig:nephonon}(b), we compare, as an example, the $z$-component 
of the spin for both models. Again, the qualitative form of the spin
expectation value as a function of the bias voltage agrees for both types of
calculations, apart from small deviations and different peak  heights. 

\section{Conclusions}
\label{sec:conclusion}
The interplay of ferromagnetism, Coulomb and electron-phonon
coupling in quantum dots or molecular transistors gives rise to interesting
spintronic effects under nonequilibrium conditions. In the
presence of ferromagnetic leads, the nonmagnetic molecular bridge supports a
coherent time evolution of the spin between the $|\up\rangle$ and
$|\down\rangle$ states.
When the tunnel coupling is weak, we obtain the stationary density
matrix in the presence of a finite bias voltage by means of the real-time
diagrammatic technique. Thereby, we include all quantum coherences that
govern the coherent evolution of the confined electron spin. As a central
result, we find a phonon-induced exchange magnetic field which acts on the
electron spins confined in the intrinsically nonmagnetic dot. The
electron-phonon coupling induces an effective attractive Coulomb interaction,
which, 
together with the spin-valve effect, induces a precession of the confined
electron spins in the dot. Interestingly enough, we find that the induced
exchange magnetic field points in the opposite direction when 
$\lambda\neq 0$, as compared to $U\neq 0$. The phonon-induced exchange field
shows much broader resonances than the Coulomb-induced exchange field. This
behavior is related to the
small energy spacings between the adjacent energy levels in the polaron picture.
However, once a single spin is confined on the dot, it starts to precess in
the effective exchange field. When $\lambda$ is finite, Franck-Condon physics
emerges. In particular, the Franck-Condon blockade give rise to an empty dot,
and counteracts the spin-valve physics. Nevertheless, if both interactions are
of the same order of magnitude, spin-valve physics survives, and spin
accumulation together with a spin precession is observable. These effects are
to some extent weaker as compared to the standard pure spin-valve scenario. 
Also the symmetrized tunneling current and the differential
conductance are accessible. In addition to the known results for
the conventional quantum dot spin-valve and the established Franck-Condon
sidebands, we propose a combined $dI/dV$-diagram, where the negative
differential properties of the quantum dot spin valve are modulated by the
Franck-Condon sidebands. 
These theoretically proposed effects might be
observable in ultraclean carbon nanotubes or gated single molecule  
setups, for instance, by magnetic scanning tunneling microscope
experiments in the near future.

\section*{Acknowledgments}   
We thank J\"urgen K\"onig, Philipp Stegmann, and Peter Nalbach 
for fruitful discussions. 
Financial support by the DFG Schwerpunktprogramm ``Spin Caloritronics'' 
 (SPP 1538) for JB and by the DFG Sonderforschungsbereich 668
``Magnetismus vom Einzelatom zur Nanostruktur'' is acknowledged
as well.

\end{document}